\documentstyle[12pt]{article}
\begin{document}
\textwidth 6in
\textheight 9in
\vskip 4cm
\Large
{\bf Hartree Fock and RPA studies of the Hubbard model}

\normalsize
\vspace{1 true  cm}

F. Guinea$^1$, E. Louis$^2$  and J. A. Verg\'es$^1$
\vspace{0.5 true  cm}

$^1$
Instituto de Ciencia de Materiales.
Consejo Superior de Investigaciones Cient{\'\i}ficas.
Cantoblanco. 28049 Madrid. Spain. \\

$^2$
Departamento de F{\'\i}sica Aplicada.
Universidad de Alicante.
Apartado 99. 03080 Alicante. Spain.

\vspace{.5 true  cm}



\large{\bf 1 Introduction.}
\normalsize

	The Hubbard model has been the simplest starting point in the
study of strongly correlated systems\cite{hubbard,kanamori,gutzwiller}.
Almost all methods known to condensed matter
theoreticians
have been applied to its analysis.

	Much progress has been achieved in the 1D version of the model,
starting from the exact description of the ground state
wavefunction\cite{lieb}. Its behavior as function of the dimensionless
coupling constant, $U/t$, and band filling, $n$, is representative
of all 1D Luttinger liquids. These are characterized by a vanishing
quasiparticle pole at the Fermi surface, a gapless spin
and charge excitation spectrum, and the separation of
spin and charge.

	The situation in other dimensions is much more
unclear. There is ample evidence that the model, at half
filling and in a bipartite lattice, is an insulator
with long range antiferromagnetic order. Some progress has
been made in the study of the model in
infinite dimensions\cite{infty} and at half filling,
where a metal insulator transition can be proved to exist
in any lattice. The limit of a low electron density starts
to yield to a variety of techniques\cite{randeria}. The
established consensus is that the system behaves like
a normal Fermi liquid, although the 2D case is still
open to controversy.
The rest of the phase diagram, especially in the
interesting case of 2D near half filling, remains to be
understood satisfactorily.

	Among the many methods used in the analysis of the model,
it is somewhat surprising the low attention given to
standard mean field schemes, like the Hartree Fock
approximation, and extensions thereof. Most of our
understanding of conventional materials has been obtained
from these methods. On the other hand,
the electron-electron interaction is, in general,
comparable to the other terms in the
hamiltonian.
The success of the Hartree-Fock methods
is due to the fact that most
effects of the electron-electron interaction can be
accounted for by means of an appropiately chosen effective
potential. The validity of Landau's description of quasiparticles
in a Fermi liquid depends on this assumption. A rather
comprehensive description of the properties
simple metals can be obtained from
the Hartree-Fock and Random Phase approximations.

	Despite the successes reported above, mean field
schemes are traditionally considered inadequate for the
study of strongly correlated systems. To our knwoledge, two main
reasons are argued to support this conclusion:

	- The Hartree-Fock method fails to describe,
even qualitatively, the physics of the "canonical"
model of strongly correlated electrons, the Anderson
impurity. No hint of the Kondo effect can be deduced
from such an analysis.

	- The other well understood unconventional
strongly correlated system is the 1D Luttinger
liquid. Again, a naive application of standard
mean field techniques fails to capture the main
physical properties, described before.

	It is worth to study in detail the adventages
and shortcomings of the Hartree Fock approximation in view
of these two apparent catastrophical failures:

	- HF and the Anderson model. The Anderson model
is an isolated "Hubbard atom" interacting with a Fermi
gas. Its behavior shows different regimes, characterized by
the energy scales of the problem:

	Below energies of order $U$, double occupancy of the
site becomes irrelevant, and the only degree of freedom left
is the spin of the impurity.

	There is a lower scale defined by the hopping of
the electrons to and from the impurity, $\Gamma$. At energies
where double occupancy is frozen out, this term generates
an effective antiferromagnetic coupling, $J \sim \Gamma^2 / U$.

	Finally, there is still a lower scale, which
gives the binding energy of the singlet formed
by this antiferromagnetic coupling,
$T_K \sim J \exp [ - N ( \epsilon_F ) J ]$.

	The system shows a crossover between a high temperature
regime, where the impurity spin fluctuates, weakly coupled to
the rest of the $N$ electrons, and a low $T$ behavior,
characterized by an almost frozen singlet,
decoupled from the remaining $N - 1$ electrons.

	The failure of the HF approximation to capture the
low $T$ physics can be traced back to the existence
of this crossover. A single effective potential
will never describe two quite different regimes.
The potential adjusts itself to optimize the description of most
of the effects which give shape to the wavefunction.
The processes above $T_K$ are reasonably described
by the H-F method, from the suppression of double
occupancy down to the existence of an antiferromagnetic
interaction.

	- The 1D Luttinger liquid. Here, the situation is
different, as these systems do not exhibit any kind
of crossover. Standard application of HF leads to a
spin density wave (for the 1D Hubbard, for example),
with a gap in the spectrum, for any filling.
The existence of this gap, and the breaking of translational
invariance by the SDW are taken as serious failures
of the method. However, it can be shown that
these features are not such unsurmountable problems:

	The 1D Hubbard model does not exhibit a true
gap, even within HF, except at half filling.
In order to compute the energy required to add an extra
electron, it is best to compare the total energies
of wavefunctions with different number of electrons.
These energies show no discontinuity, except at half
filling. The extra electrons do not fill the
lowest state above the gap in the initial solution,
but give rise to a rearrangement of the SDW and of the
wavefunction. Moreover, this rearrangement
of the entire many body state leads to
$Z = | < \Psi_N | {c^+}_k | \Psi_{N-1} > |^2 \rightarrow 0$.
Thus, the dissapearance of the quasiparticle pole
can also be described within HF.

	The breaking of translational symmetry is required in order
to account for the strong shake up effects induced
by extra electrons, which lead to the vanishing
of coherent quasiparticles. The random phase approximation
suffices to restore this symmetry,  giving a zero
mode associated with the motion of the SDW. The RPA also
gives a difference between the velocities
of the low energy spin and charge collective excitations,
which are missing at the HF level.

	Thus, a judicious application
of the HF and RPA methods leads to the most relevant
features of the 1D Luttinger liquid: absence of the quasiparticle
pole, gapless excitations, and separation of spin and
charge.

	In the following, we will make use of these
considerations to explore the phase diagram of the 2D
Hubbard model in a square lattice.
\vskip 0.5cm
\large{\bf 2  Mean field solutions.}
\normalsize

	The Hartree Fock scheme can be implemented
with a variety of contraints. Usually, the equations
are solved in Fourier space, assuming that some kind
of spin and charge density wave exists. The
possible wavevectors for these distortions are given by
the size of the unit cell considered. The amplitudes
of these possible waves can be seen as variational parameters
to be optimized.

	This procedure suffices to give a reasonable description
of a conventional system in most cases. For the Hubbard model,
and at half filling, we find as the best solution
an antiferromagnetic SDW, with the spins
in the two sublattices pointing in opposite directions.
A normal, paramagnetic metal is also described by a Slater
determinant made up of plane waves, which is
a HF solution of the Hubbard model with a sufficiently
dilute concentration of electrons.

	Near half filling, however, many solutions
appear which cannot be described in such a simple
way. Thus, we are forced to solve the
selfconsistent HF equations in finite clusters, without
imposing any constraints on the shape of the
effective potential. Open and periodic boundary conditions
are used. The most striking result obtained is that
{\it many different selfconsistent solutions (local minima)
are found}. As the solution is found by an iterative procedure,
the final shape depends on the initial guess for the
wavefunction. This fact is presented in fig. (1),
where two typical textures are shown (for a
more detailed description of the solutions, see\cite{HF}).

\vskip 10cm

\small{Figure 1. HF antiferromagnetic ground state at half filling
(left), and spin and charge texture for 10\% doping (right).
The arrows denote the spins, ant the circles stand for
the deviation of the local charge from 1 (half filled band).}
\normalsize
\vskip 0.5cm
Clearly, these wavefunctions break translational
symmetry, as in the 1D case mentioned in the introduction.
The number of solutions greatly increases with the size
of the system. Note that the cluster sizes are much larger
than the typical correlation length in the AF phase,
$\sim t / \Delta$ lattice units, where $\Delta$ is
the AF gap. The energies of these textures
fluctuate strongly with cluster size, filling
and value of $U/t$. Thus, it is impossible,
in practical terms, to define the best HF solution
of a cluster of moderate size, near half filling.

	This richness of textures difficults the
study of the model, but also points out to
important physical features of the Hubbard model.
The large number of metastable minima of the
HF energy suggest the existence of glassy
behavior. The underlying frustration is due
to the competition between the repulsion and
the kinetic energy, complicated further
by anisotropy effects, which are greatly
enhanced near half filling.

	The solutions can be classified, in broad terms,
into {\it magnetic polarons}, {\it domain walls} and
{\it vortices}. Domain walls are a direct consequence of the
nesting near half filling\cite{schultz}, and are most stable at low
doping and small values of $U/t$. Magnetic polarons are localized
textures which resemble small ferromagnetic bubbles in an
AF background. They dominate for large values of $U/t$.
There is a smooth crossover from extended polarons to
circular domain walls, as function of doping and
$U/t$. Finally, vortices arise because of the additional
$U(1)$ symmetry provided by the phase of the wavefunction.
They tend to appear at intermediate values of $U/t$.
Periodic arrays of these solutions are found for
sufficiently large cluster sizes. There is a wide variety
of other, less regular, textures in large clusters.
Among them, it is worth mentioning that, for certain dopings,
holes tend to form localized pairs with p-wave symmetry. These
pairs have a finite binding energy. These solutions suggest
the possibility of superconductivity for a small
region of parameter space.
\vskip 0.5cm

\large{\bf 3  Excitation spectrum. RPA.}
\normalsize

	We have computed the excitations within the Random Phase
Approx- \hfil
imation\cite{RPA}. This extension of the Hartree Fock method is
crucial in order to obtain correctly the low energy spin waves.
The initial HF excited states, however, are a reasonable
guide for the charge excitations, which are not so
strongly modified  by the inclusion of the self consistency
by the RPA.

\vskip 10cm
\small{Figure 2. Density of states for the two HF solutions shown in
figure 1. The chemical potential for the doped case lies at
$\omega = 0.95$.}
\normalsize
\vskip 0.5cm
	The HF density of states is shown in  fig. (2), for two
characteristic situations: a) At half filling, the
system is an insulator, whith a gap in the charge excitation
spectrum of order $U$. b) Near half filling, a number of
localized states appear within the AF gap. The Fermi level
remains at the edge of the relevant band. The level
structure is by no means rigid, as the number of
states within the gap is proportional to the number of
holes.
Note that all
these features are consistent with results
obtained from the exact diagonalization of
finite clusters\cite{exact}.

	The states within the gap imply that, away from
half filling, a continuum of low energy charge excitations
exists. This picture for the charge-charge channel
remains qualitatively unaltered when the RPA is fully
implemented. On the other hand, the spin waves can
only be described within the RPA. As function of
doping, the total spectral strength in the spin wave
band decreases proportionally to the number of electrons or holes.
The density of low energy excitations,
for a case near half filling, is shown in fig. (3).

	We have also computed, because of its experimental
relevance, the density of excitations at the corner
of the Brillouin zone, ( $\pi , \pi$ ). At half filling, there
is a large peak at zero energy, due to the AF ordering.
This peak is broadened, and shifts to higher energies,
as the doping increases. There is no significant structure
in the charge spectrum. This difference in the \hfil
spectral
weights throughout the Brillouin Zone between the charge
and spin excitations persists up to quite large
doping levels ( $\approx 50$ \%, for $U/t = 10$ ).

\vskip 10cm
\small{
Figure 3. Density of spin and charge excitations for the
two HF solutions shown in figure 1. Dotted lines give the
spin and dashed lines the charge excitations. Thick lines
represent results for the doped cluster. The curves in the left
side are integrated densities of excitations, while those
at the right are excitations at the $( \pi , \pi )$
point of the Brilouin zone.}
\vskip 0.5cm
\normalsize
The fact that charge and spin excitations show different
dispersion in momentum space is an indication of
spin-charge separation. As mentioned in the
introduction, the RPA suffices for a qualitative description
of this phenomenon in 1D. In our case, the
existence of a 2D dispersion relation,
instead of simple sound waves, allows for a richer
variety of phenomena, which will be studied elsewhere.

\vskip 0.5cm
\large{\bf 4  Quasiparticle strengths.}
\normalsize

	The proliferation of HF solutions which break translation
symmetry suggests the relevance of shake up effects, induced by
the injection of electrons or holes. As mentioned in the
preceeding section, the HF energy levels do not shift rigidly
as function of the electronic density. The textures
rearrange themselves, so as to accomodate the electrons
in newly formed localized levels. In turn, these levels
modify the underlying continuum.

	The importance of these shake up effects can be
estimated, within the Hartree Fock approximation,
by computing the matrix element: $Z = | < \Psi_0 ( N ) |
c_i | \Psi_0 ( N + 1 ) > |^2$. The operator $c_i$ is taken to be
the lowest occupied electronic level in $| \Psi_0 ( N + 1 ) >$.
This procedure, although
unusual in applications of the HF method, is
perfectly well defined as an approximation to the
spectral strength of the quasiparticle pole.
In normal metals, this scheme gives $Z = 1$. This is due to
the fact that most of the missing spectral strength goes into
plasmon satellites. The description of this structure
requires us to go one step beyond
the HF + RPA procedure used here. In the
Hubbard model, the depletion of
the quasiparticle pole comes from low energy, electron-hole pairs.
Their shake-up is approximately described by the rearrangement
of the electronic levels in the HF Slater determinant.
Thus, we expect, at least, a qualitative description of this effect
within our approximations.

	The results that we find are discussed in detail
elsewhere in this volume (see also\cite{QP}). The most relevant feature
is that, near half filling, the splitting from the continuum
of localized states upon doping, leads to a vanishing $Z$.
This result can be viewed as an "orthogonality catastrophe",
induced by the phase shifts generated by these
localized states in the continuum levels. Understood in these terms, the
nature of the physical regime that we find near half doping
is very similar to the 1D Luttinger liquids.

\vskip 0.5cm
\large{\bf 5   Conclusions.}
\normalsize

	We have presented a comprehensive study of the two
dimensional Hubbard model in a square lattice, using mean field
techniques. We find three different regimes:

	- The system is an antiferromagnetic insulator
at half doping. The low energy excitations are spin waves.

	- Very far from half doping, the model
behaves as a conventional Fermi liquid. The low energy
excitations are electron-hole pairs. The doping required
to reach this regime depends on the value of $U/t$, but
it is always present.

	- Near half doping, we find a phase quite reminiscent
of the 1D Luttinger liquids. The system has no gap.
At low energies, spin and charge excitations coexist,
although with different dispersion relations.
There are strong shake-up effects upon the injection of
electrons and holes, which lead to a vanishing quasiparticle
pole.

	In addition, our scheme also gives a
ferromagnetic phase\cite{nagaoka},
for large values of $U/t$, and very close to half doping.
This phase remains to be fully characterized.

\end{document}